# STATIC MAGNETIZATION PROPERTIES OF AL800 GARNET MATERIAL

J. Kuharik[†], R. Madrak, A. Makarov, W. Pellico, S. Sun, C. Y. Tan, and I. Terechkine,
Fermi National Accelerator Laboratory, Batavia, IL, USA

*Abstract*

A second harmonic tunable RF cavity is being developed for the Fermilab Booster. This device, which promises reduction of the particle beam loss at the injection, transition, and extraction stages, employs perpendicularly biased garnet material for frequency tuning. The required range of the tuning is significantly wider than in previously built and tested tunable RF devices. As a result, the magnetic field in the garnet comes fairly close to the gyromagnetic resonance line at the lower end of the frequency range. The chosen design concept of a tuner for the cavity cannot ensure uniform magnetic field in the garnet material; thus, it is important to know the static magnetic properties of the material to avoid significant increase in the local RF loss power density. This report summarizes studies performed at Fermilab to understand variations in the magnetic properties of the AL800 garnet material used to build the tuner of the cavity.

## INTRODUCTION

In the Booster accelerator at Fermilab, the frequency of the accelerating RF cavities changes during the cycle from 37.77 MHz at injection to 52.81 MHz at extraction. Frequency tuning of these cavities is accomplished by using longitudinally biased ferrite as described in [1]. If second harmonic cavities are added to the accelerating system, injection and extraction efficiencies can be noticeably improved. The problem is that with twice as wide required frequency range, power loss in the tuners of the cavities becomes prohibitively high if longitudinally biased ferrite is used. Using instead perpendicularly biased garnet material promises significant reduction in the RF power loss at the expense of an increased bias field. Prototypes of these kinds of devices were built and tested [2-3], but never used in accelerators because the tests at high power resulted in overheating in the garnet rings. Analysis of the magnetic system of these devices brought us to the conclusion that the reason for the overheating could be anomalously low magnetic field in the garnet that filled the cavities' tuners. This low field level, which can come close to the gyromagnetic resonance, results from the strong nonlinear behaviour of the garnet material. To properly model the cavity during the design stage, which should include expected temperature of the garnet in the tuner, knowledge of static magnetic properties of AL800 material was imperative. Corresponding studies were initiated at Fermilab; the main results are presented in this report.

As fabrication of the garnet rings for use in the tuner is a complicated multi-stage process, it was also considered important to understand the range of the magnetic properties expected from different batches. Large differences would result in additional uncertainties due to azimuthal non-uniformity of the RF losses in the tuner.

## AL-800 GARNET PERMEABILITY: FIRST ITERATION

The first garnet permeability measurements were made using available sample rings of AL800 material. The rings were placed inside an existing magnetic system (a solenoid and a flux return) that generated the bias magnetic field [4]. At each setting of the bias current in the solenoid, the magnetic field was measured at several locations along the stack of ten rings using cross-calibrated Hall probes. The magnetic field measured at these locations was compared with results of thorough magnetic modelling that assumed non-linear magnetization properties of the garnet. Starting with an initial magnetization curve, it was iteratively adjusted to ensure that the modelling properly reflected the data obtained by direct measurements within the total range of used bias currents. Fig. 1 shows the static magnetic properties of AL-800 garnet obtained as a result of this procedure.

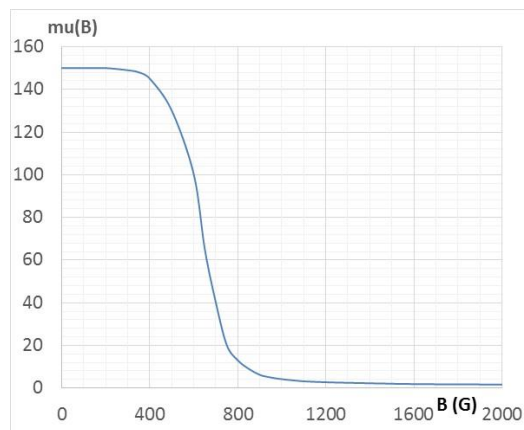

Figure 1: Static permeability of AL-800 garnet measured using existing garnet rings.

Significant non-linearity of the curve in the region around the saturation magnetization (767 Gauss for the used material) results in a substantial spatial change of the magnetic field in the garnet if the bias field is not perfectly uniform. Although some measures may be taken to mitigate this effect (e.g. see [5]), results can differ depending on the magnetic properties of garnet used to fabricate rings for each cavity tuner.

This manuscript has been authored by Fermi Research Alliance, LLC under Contract No. DE-AC02-07CH11359 with the U.S. Department of Energy, Office of Science, Office of High Energy Physics. The U.S. Government retains and the publisher, by accepting the article for publication, acknowledges that the U.S. Government retains a non-exclusive, paid-up, irrevocable, world-wide license to publish or reproduce the published form of this manuscript, or allow others to do so, for U.S. Government purposes.

Due to the limitations imposed by technological process at the vendor's (National Magnetics Group, Inc.) site, each garnet ring in the tuner of the prototype second harmonic cavity will be assembled from eight sectors. Some spread of magnetic properties is inevitable because of deviations in the proportions of the mixed materials and spatial non-uniformity of temperature in the sintering kilns. To understand the size of the spread, two "witness" samples were cut from each sector and their magnetic properties were measured.

## WITNESS SAMPLES PERMEABILITY MEASUREMENT

The samples are cylinders 17.8 mm in diameter and 16 mm long. To make sure the magnetic measurements on the samples give reliable results, the magnetic field in the samples must be close to uniform at every setting of the bias current. The small size of the samples presented some challenges to this design goal. The design concept of the measurement fixture is shown in Fig. 2, where 1/8 of the full setup is shown.

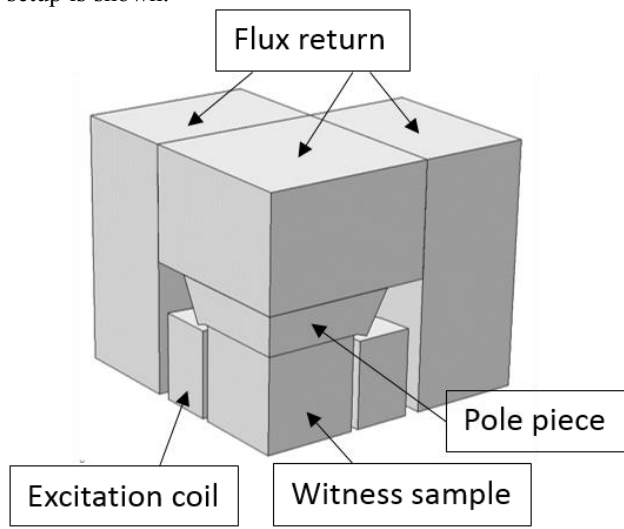

Figure 2: Witness sample measurement setup.

The material of the flux return is MN-60 ferrite with an initial permeability ~6000. Profiled pole tips of the flux return ensure better than 4% uniformity of the permeability in the samples when the magnetic properties are those shown in Fig. 1 and the range of the bias field is from 0 to ~3000 G. No saturation of the flux return is expected.

Two coils are used during the measurements: the excitation coil, which sets the magnetic field, and the signal coil, which is used to extract the information about the average magnetic field in the sample during the measurements. The number of turns in the excitation coil is 183; it is wound using #14 (Ø1.62 mm) insulated copper wire. The number of turns in the signal coil is 80; #34 (Ø0.16 mm) insulated copper wire was used for this winding, which was placed close to the surface of the sample.

The sample measurement circuit is shown in Fig. 3. The bias AC field is generated by an Agilent 33250A signal generator and amplified by the Behringer iNuke 6000 amplifier, which can generate currents up to ~20 A in the excitation coil. This current is measured using a precisely known shunt resistance (R = 3.92 Ω). A 1000 µF blocking capacitor prevents the DC component of the current from biasing the magnetic circuit. It was found that this DC bias, if not removed, could result in substantial distortion of the measurement results. The excitation current and the voltage generated in the signal coil are recorded using a Tektronix TDS5054B-NV digital oscilloscope.

Before each measurement, the sample was demagnetized using a decaying sine wave with the initial amplitude of the current 25% greater than the maximum current used in the previous measurement cycle. After each measurement, the apparatus was taken apart and the sample was inverted for the next measurement to avoid any remnant magnetization or other systematic errors.

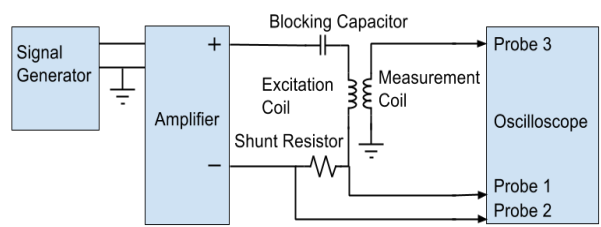

Figure 3: Witness sample measurement block diagram.

The excitation current during the measurement cycle is a sine half-wave with a peak current up to 20 A. During the measurements, voltage to ground is measured at either side of the shunt resistor (probes 1 and 2) and across the measurement coil (probe 3). The difference between the voltages on probes 1 and 2 is used to calculate the current and magnetic field in the sample (as the magnetic properties of the flux return in the sample holder are known). The measured voltage across the signal coil allows the calculation of the magnetic flux through the sample by numerical integration. Two additional corrections must be made to achieve acceptable reproducibility and accuracy of the magnetization curve measurement. First, the internal DC offset of the scope must also be accounted for in the calculations. If not compensated for, this offset introduces noticeable systematic error in the results of numerical integration. Second, a small (~0.25 mm) air gap between the witness sample and the signal coil must be also accounted for in the calculations. This air gap contains magnetic flux that contributes to the measured voltage, but must be excluded when the magnetic flux in the sample is calculated.

The first set of measurements was made on 16 samples of garnet material cut from the slabs used to fabricate the first garnet ring. The samples were measured in the high field region, which is the working point of the garnet material in the tuner. In this region, the magnetic flux density in the garnet is well above the gyromagnetic resonance. In Fig. 4 typical traces registered during the measurements are shown. Digital records of these traces are corrected as mentioned above and the permeability data is extracted.

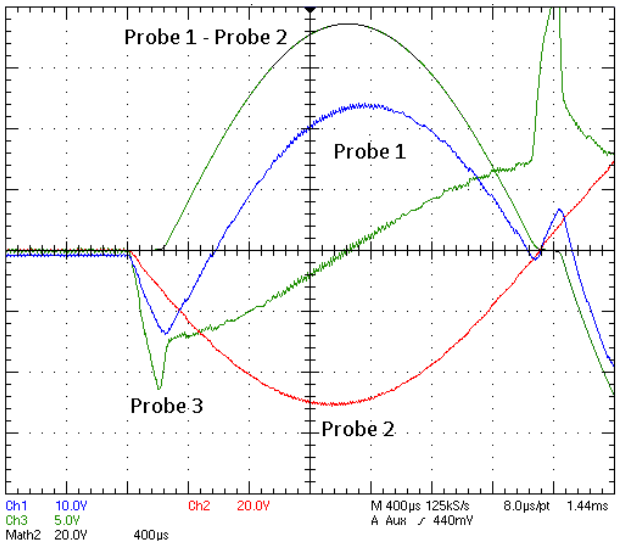

Figure 4: Typical scope trace. Probes #1 and #2 are placed on either end of the shunt resistor; probe #3 measures the output of the signal coil.

Results of the witness samples magnetization measurements for the bias magnetic flux density $B > \mu_0 M_S$, where $M_S$ is saturation magnetization of the material, are summarized in Fig. 5. In this figure, the average values of the material permeability are represented by a solid curve. The graph also shows:

- The theoretical RF permeability of the garnet material calculated using Ref. [6] in the high bias field region: $\mu(B) = B/(B-\mu_0 M_S)$.
- The permeability data for the high field region measured in [4] and extracted from the full set of data shown in Fig. 1.
- Spread in the calculated permeability is shown by a diffused area in the plot.

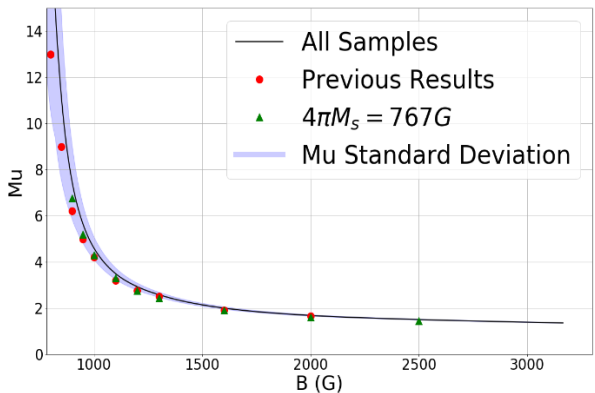

Figure 5: Summary of the measurement results in the high field region.

As one can see, the static permeability measured in the high field region basically follows the RF permeability. This is not the case for the measurements made in the low bias field region. Knowing the behaviour of the static permeability in the low field region is important when the bias field in the tuner of the cavity is generated using only one winding and hence only one (pulsed) power supply. In this case, the power supply drive voltage depends not only on the current rise rate, but also on the change of the inductance in the magnetic circuit of the bias system, which is a function of the static permeability of the garnet material.

Because of the large range of the expected permeability change (Fig. 1) and imperfect flatness of the pole pieces of the flux return, current permeability measurements in the low bias field region are not very reproducible. However, they do confirm that the maximum permeability is at least as large as shown in Fig. 1. Some changes in the mechanical design of the pole tips of the flux return are being implemented in order to improve measurement reproducibility.

Apart from the measurements made using the witness samples, assembled garnet rings will also be measured in a specially designed test cavity. Each garnet ring procured from the National Magnetics Group, Inc. will be placed between the poles of a magnet generating the perpendicular bias field in the tuner of a test cavity. The entire frequency range (75 – 110 MHz) is expected to be covered. Measurements in this setup will provide data about the average permeability and the loss tangent of the material in each garnet ring as function of the bias magnetic field. Conceptual sketch of the ring permeability measurement setup is shown in Fig. 6.

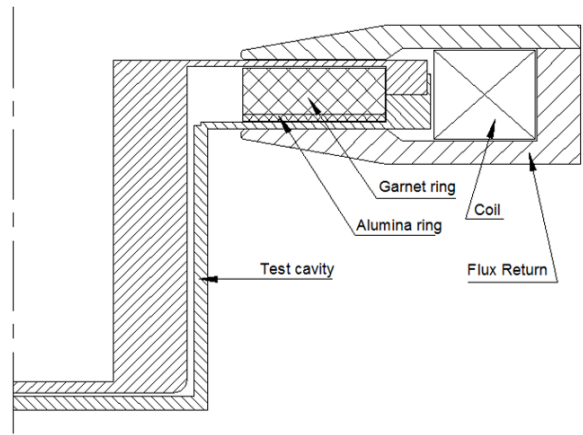

Figure 6: Ring test setup.

## SUMMARY

A systematic study of the magnetic behaviour of AL800 garnet material has been conducted to support the ongoing efforts to finalize the design of a second harmonic cavity for the Fermilab Booster. Witness samples are cut from each piece of the material used to assemble the ring-shaped blocks of garnet in the cavity tuner. Their magnetic properties are measured to verify the uniformity of the permeability in the garnet sectors used to assemble each block. The measurements of the samples associated with the two first procured garnet rings made in the high magnetic field region show satisfactory uniformity. Low field measurements still require refinement of the measurement technique and the setup to get reliable results.